\documentclass[11pt,twoside]{article}

\usepackage{asp2006}
\usepackage{epsf}
\usepackage{psfig}
\usepackage{lscape}

\begin{document}
\title{3D Radiative Transfer Modeling of Clumpy Dust Tori Around AGN}   %%% Fill in title
\author{S.~F. H\"onig, T. Beckert, K. Ohnaka, G. Weigelt}   %%% Fill in author names
\affil{Max-Planck-Institut f\"ur Radioastronomie, Auf dem H\"ugel 69, 53121 Bonn, Germany}    %%% Fill in author affiliations

\begin{abstract} %%% Abstract to run on from here.
We present 3-dimensional radiative transfer models for clumpy dust tori around AGN. Our method combines
Monte Carlo simulations of individual dust clouds with the actual 3-dimensional distribution of clouds
in the torus. The model has been applied to NIR and MIR photometric and interferometric observations of
NGC~1068. For the first time, it is possible to simultaneously reproduce both photometric and interferometric
observations in the NIR and MIR. We infer a luminosity $L=2\times10^{45}\,{\rm erg/s}$ and an inclination of $i=70\deg$ for
NGC~1068 from our model.
\end{abstract}

%%% MAIN BODY OF TEXT GOES HERE. CONSULT "INSTRUCTIONS FOR AUTHORS USING
%%% LATEX2E MARKUP", SECTIONS 2.3-2.6 FOR HELP WITH EQUATIONS, FIGURES,
%%% AND TABLES.

\section{Introduction}   %%% Top level section head (remove "%" symbol)
The presence of an optically thick dust torus is the cornerstone of unification schemes for AGN to explain the
orientation-dependent difference between type 1 and type 2 objects. The dust inside the torus absorbes the AGN radiation
and reemits it in the infrared. For the nucleus of NGC~1068, interferometric observations directly resolved the torus in the NIR and MIR
\citep[e.g.,][]{Wei04,Jaf04}. The interferometric data support the idea that the dust within the torus is concentrated in clumps
instead of homogeneously distributed \citep{Wit04,Jaf04}.

\section{Clumpy torus model and its application to NGC~1068}
\begin{figure}[!ht]
\plottwo{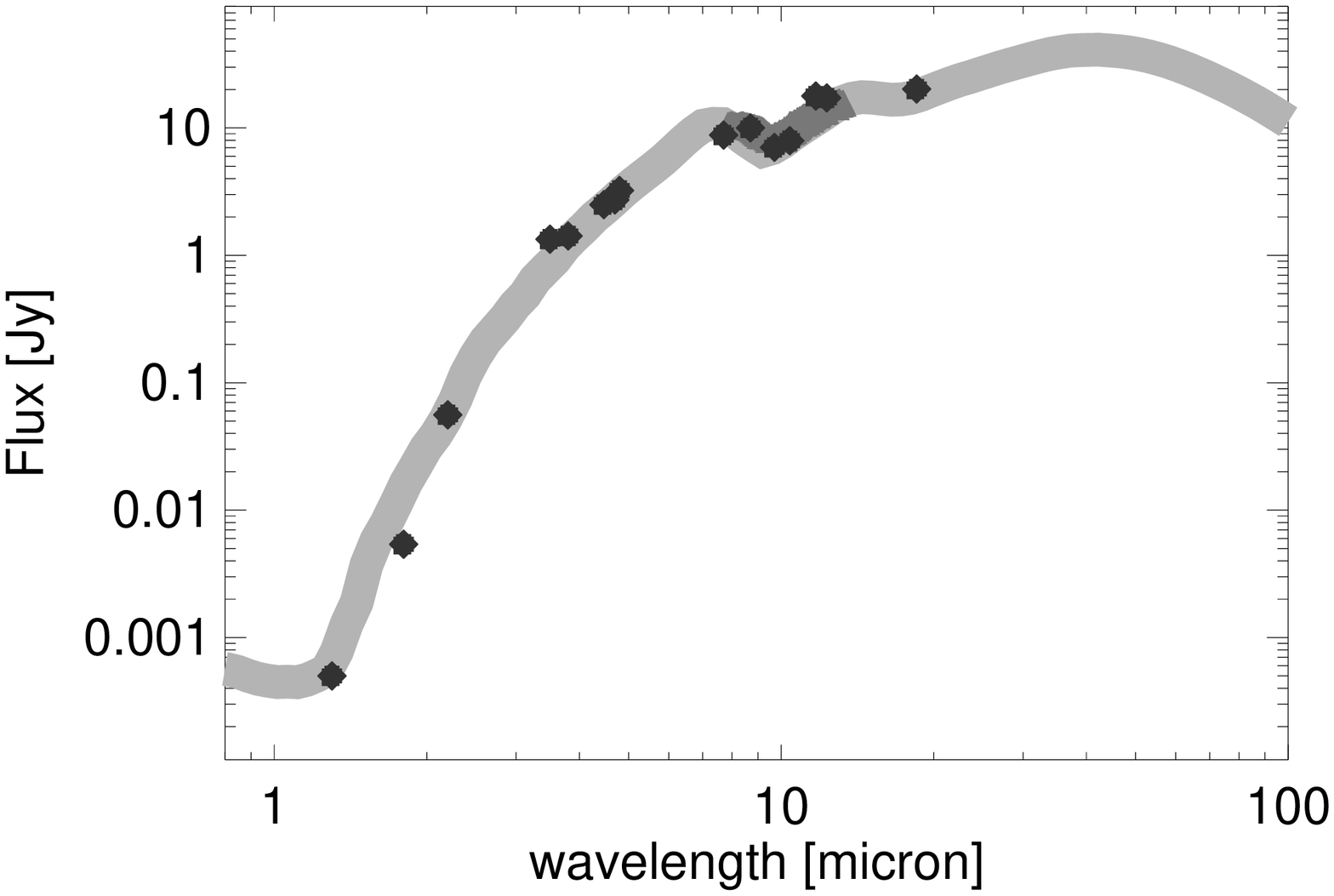}{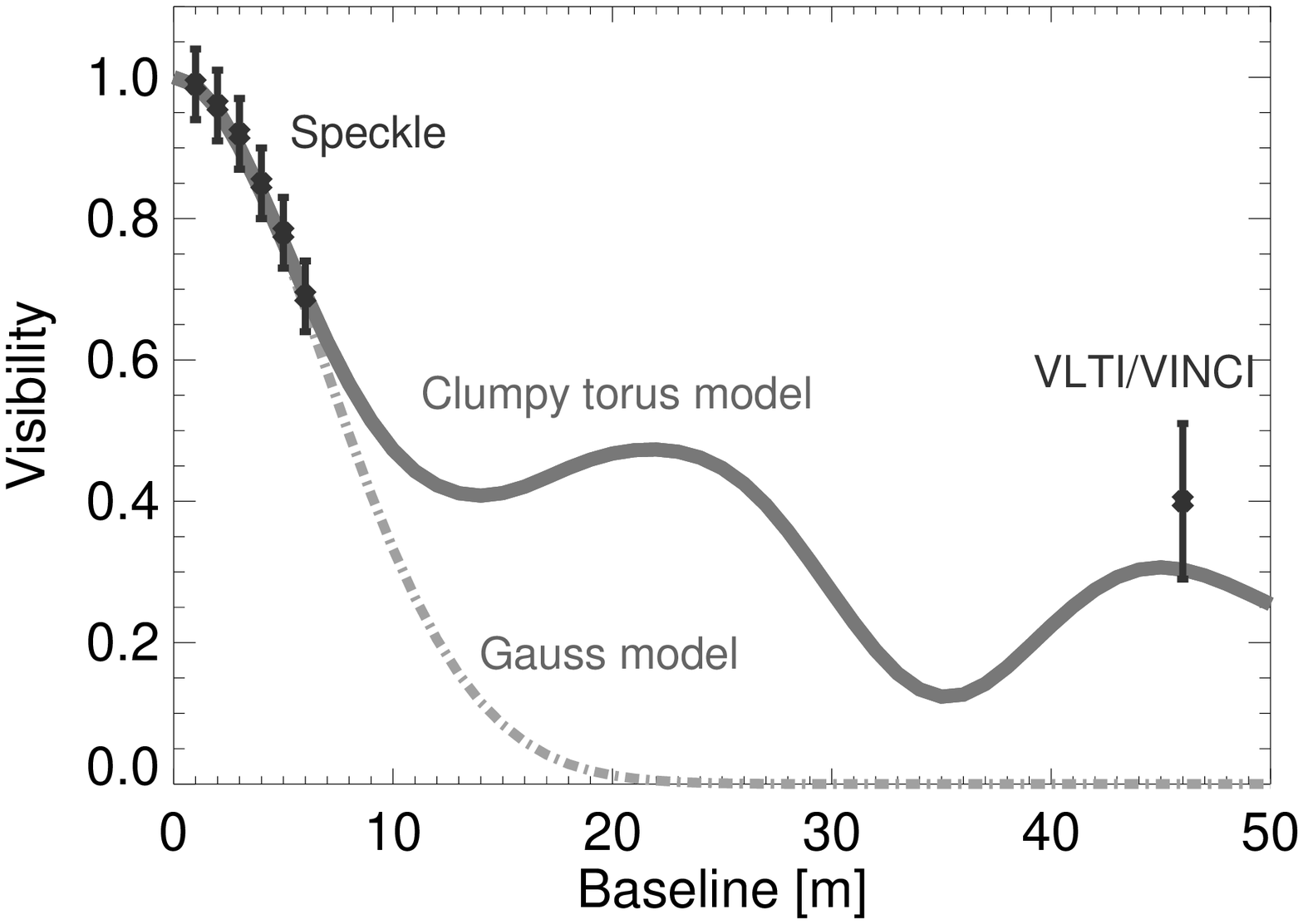}
\caption{Comparison of our clumpy torus model with data for NGC~1068. {\it Left:} Observed (dark marks) and model SED of NGC 1068.
{\it Right:} $K$-band interferometry and model visibility.}\label{N1068SED}
\end{figure}

Recently, we presented our new 3-dimensional radiative transfer calculations of clumpy dust tori \citep[for more details, see][]{Hon06}.
The torus modeling makes use of a database of dust cloud SEDs simulated with our Monte Carlo radiative transfer code. For each set
of model parameters, we simulate torus SEDs for different random arrangement of clouds. Our
torus model was applied to the Seyfert 2 AGN NGC~1068 \citep{Hon06}. Here, we present new results with some improvement
to the earlier publication. We simultaneously modeled NIR and MIR photometry as well as NIR and MIR interferometric data of the
nucleus. In Fig.~\ref{N1068SED} we show a comparison between the observed high-resolution SED and our model SED. Important model
parameters are the AGN luminosity $L_{\rm bol}=2\times10^{45}\,{\rm erg/s}$, torus inclination $i=70\deg$, and optical depth of the
torus $\tau_{\rm Torus}(\lambda=10\,{\rm \mu m})\sim 1$.

\begin{figure}[!ht]
\plotone{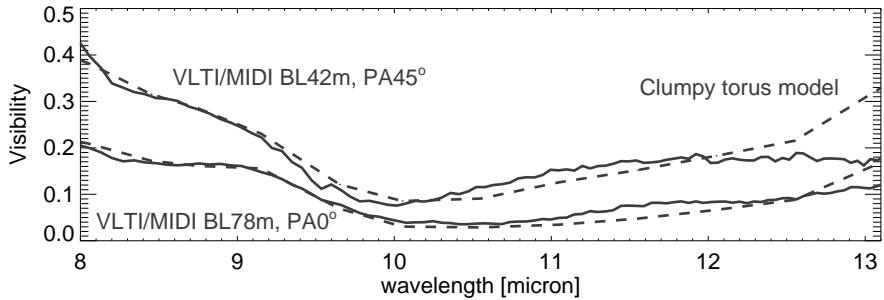}
\caption{Comparison between the VLTI/MIDI spectro-interferometry of NGC 1068 for two different baselines (solid lines) and our clumpy torus model
(dashed lines).}\label{N1068MI}
\end{figure}

Why do we need clumpiness in the torus? \citet{Kro88} theoretically argue that homogeneously distributed dust cannot
survive in the environment of an AGN. This has been supported by MIR spectro-interferometric observations of NGC~1068 which find deeper silicate features
with growing baseline \citep{Jaf04}. This is apparently inconsistent with homogeneously distributed dust \citep{Sch05}. In addition, NIR long-baseline interferometry
of NGC~1068 finds a surprisingly high visibility at a baseline of 46\,m which cannot be explained by a simple homogeneous model \citep{Wit04}.
In Figs.~\ref{N1068SED} \& \ref{N1068MI}, we show that our clumpy torus model is able to reproduce the interferometric results.

%\begin{figure}[!ht]
%\hspace{-22cm}
%\plotfiddle{n1068c.eps}{4.8cm}{0}{17}{17}{1}{1}
%\caption{Comparison between the observed high-resolution SED of NGC 1068 and our model SED. Red symbols: high-resolution
% photomety; Green \& blue curves: MIR spectra \citep{Jaf04,Mas06}. The shaded area shows the range of model SED variations obtained for 10
% different random cloud arrangements. The dark grey curve shows the best-fitting arrangement.}\label{N1068SED}
%\end{figure}

%\begin{figure}[!ht]
%\plotone{MIDIvis.eps}
%\caption{Comparison between the VLTI/MIDI spectro-interferometry of NGC 1068 (Jaffe et al. 2004) and our clumpy torus model.
% Red \& green curves show the observed visibilities, blue-dashed lines are predictions from our best-fitting model
% (see Fig.~\ref{N1068SED}).}\label{N1068MI}
%\end{figure}
%
%\begin{figure}[!ht]
%\plotone{visiBL.eps}
%\caption{$K$-band interferometry \citep{Wei04,Wit04} of NGC\,1068, in comparison to our clumpy torus model (10 different random cloud
% arrangements). The dark grey curve is the best-fitting model from Fig.~\ref{N1068SED}.}\label{N1068NI}
%\end{figure}


% \acknowledgements %%% Text of acknowledgements runs on after this command.

%%% THE BIBLIOGRAPHY
%%%
%%% CONSULT SECTION 3 OF "INSTRUCTIONS FOR AUTHORS" FOR HOW TO USE NATBIB.
%%% AUTHORS ARE ENCOURAGED TO USE EITHER THE "THEBIBLIOGRAPY" ENVIRONMENT
%%% BY UNCOMMENTING (DELETING THE "%" SYMBOL) THE COMMANDS BELOW, OR BY
%%% USING THE BIBTEX ENVIRONMENT. TO FIND OUT WHICH IS APPLICABLE TO YOUR
%%% CONTRIBUTION, CONSULT THE VOLUME EDITORS FOR YOUR PROCEEDINGS.
%%%


\begin{thebibliography}{}
\bibitem[\protect\citeauthoryear{H\"onig et al.}{2006}]{Hon06} H\"onig, S.~F., Beckert, T., Ohnaka, K., \& Weigelt, G. 2006, A\&A, 452, 459
\bibitem[\protect\citeauthoryear{Jaffe et al.}{2004}]{Jaf04} Jaffe, W. et al. 2004, Nature, 429, 47
\bibitem[\protect\citeauthoryear{Krolik \& Begelman}{1988}]{Kro88} Krolik, J. H. \& Begelman, M. C. 1988, ApJ, 329, 702
\bibitem[\protect\citeauthoryear{Schartmann et al.}{2005}]{Sch05} Schartmann, M., et al. 2005, A\&A, 437, 861
\bibitem[\protect\citeauthoryear{Weigelt et al.}{2004}]{Wei04} Weigelt, G. et al. 2004, A\&A, 425, 77
\bibitem[\protect\citeauthoryear{Wittkowski et al.}{2004}]{Wit04} Wittkowski et al. 2004, A\&A, 418, L39
\end{thebibliography}
\end{document}